\documentclass{article}

\usepackage{PRIMEarxiv}

\usepackage[utf8]{inputenc} % allow utf-8 input
\usepackage[T1]{fontenc}    % use 8-bit T1 fonts
\usepackage{hyperref}       % hyperlinks
\usepackage{url}            % simple URL typesetting
\usepackage{booktabs}       % professional-quality tables
\usepackage{amsfonts}       % blackboard math symbols
\usepackage{nicefrac}       % compact symbols for 1/2, etc.
\usepackage{microtype}      % microtypography
\usepackage{lipsum}
\usepackage{fancyhdr}       % header
\usepackage{graphicx}       % graphics
\graphicspath{{media/}}     % organize your images and other figures under media/ folder

\usepackage[figuresright]{rotating}
\usepackage{pdflscape}
\usepackage{longtable}

\title{Leveraging Artificial Intelligence Techniques for Smart Palm Tree Detection: A Decade Systematic Review 
}

\author{ {\hspace{1mm}Yosra Hajjaji}\\
	University of Manouba, Manouba, Tunisia \\
	\texttt{hajjajiyosra05@gmail.com} 
\AND
	{\hspace{1mm} Wadii Boulila}	 \\
	Prince Sultan University, Riyadh, Saudi Arabia\\
	University of Manouba, Manouba, Tunisia \\
	\texttt{wboulila@psu.edu.sa} 
\AND
   {\hspace{1mm} Imed Riadh Farah}  \\
	University of Manouba, Manouba, Tunisia \\
	\texttt{imedriadh.farah@isamm.uma.tn} 
}

\begin{document}
\maketitle

\begin{abstract}
Over the past few years, total financial investment in the agricultural sector has increased substantially. Palm tree is important for many countries’ economies, particularly in northern Africa and the Middle East. Monitoring in terms of detection and counting palm trees provides useful information for various stakeholders; it helps in yield estimation and examination to ensure better crop quality and prevent pests, diseases, better irrigation, and other potential threats. Despite their importance, this information is still challenging to obtain. This study systematically reviews research articles between 2011 and 2021 on artificial intelligence (AI) technology for smart palm tree detection. A systematic review (SR) was performed using the PRISMA approach based on a four-stage selection process. Twenty-two articles were included for the synthesis activity reached from the search strategy alongside the inclusion criteria in order to answer to two main research questions. The study’s findings reveal patterns, relationships, networks, and trends in applying artificial intelligence in palm tree detection over the last decade. Despite the good results in most of the studies, the effective and efficient management of large-scale palm plantations is still a challenge. In addition, countries whose economies strongly related to intelligent palm services, especially in North Africa, should give more attention to this kind of study. The results of this research could benefit both the research community and stakeholders. 
\end{abstract}

% keywords can be removed

\section{Introduction}
\label{main}
\subsection{Background}

\begin{figure} 
\centering
\includegraphics[width=15cm, height=8.5cm]{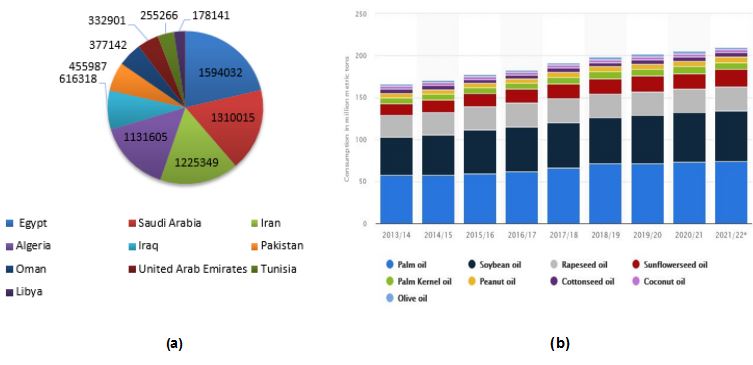}
\caption{(a) Worldwide consumption of vegetable oil by type (in a million metric tons), (b) Date production by the top ten date fruits producing countries, 2019} 
\label{fig1}
\end{figure}

The date palm (known as Phoenix dactylifera L.) is one of the most primary persisting fruit trees. It has been the most cultured fruit tree from which man has derived benefits because of its historical connection with human life sustainability. Among the 120 million predicted quantity of palm trees over the World, 70\% of the date palm are located in Arabian countries, which are at the helm for 67\% of the overall date production, according to the Food and Agriculture Organisation of the United Nations (FAO). The date palm retains its value for the farmer and his animals and in many commercial and medicinal applications.
Palm trees play a considerable role in the agricultural economy of utmost the countries in North Africa and the Middle East \cite{EI-Mously}. The top 10 larger Date Producers over the World are shown in figure.\ref{fig1}.(a).
On the other side, the oil palm (called Elaeis guineensis) is one of the ultimate vital economic crops in the World with significant economic value. Currently, oil palm is the World's largest source of vegetable oil, among other oilseeds (like rapeseed, soybeans, and sunflower) accounting for 35\% of total vegetable oil consumption \cite{Chong2017} \cite{URL1} (see figure.\ref{fig1}.(b)).
A large number of land turned into oil palm plantations, due to the increasing demand for vegetable oil, including existing arable land, especially in Indonesia and Malaysia. However, monitoring the oil palm in the planting area is helpful for farmers and planting stakeholders to increase the productivity of the planting \cite{Chong2017}.

\subsection{Palm trees Deterioration: Indicators and Causes}
Unfortunately, date palms or oil palm trees are threatened by pests and diseases  (i.e., Red palm weevil, blight spots, and beyond, the major pests in North Africa \cite{Loutfy}, socio-economic factors, and environmental changes. One of the most critical pests of palm trees is the red palm weevil, which causes much damage to palm trees and can devastate large areas of palm trees. In ten years, the productivity of date palms has declined in traditional cultivation areas. 
Managers must be provided with updated information about palm tree status. This information helps in sustainable management \cite{Fitzherbert2008}. Many planting companies practice counting trees to obtain accurate information about precision agriculture's number, distribution, and essential parts. However, it is not always possible to gather exact statistics \cite{Wulder2000}. Thus, combining modern technology and agriculture is crucial for effectively managing oil palm trees.

\subsection{Existing solutions for Palm Trees}
In the past, the most frequently used method of tree counting was done manually in the field. These approaches lead to inaccurate results. Therefore, daily using automatic methods is becoming increasingly in demand, as they represent a potentially impressive alternative for private and public agricultural institutions. In recent years, due to the rapid progress of high-resolution satellites and unmanned aerial vehicles (UAVs), automatic detection of tree crowns from remote-detected images at high resolution has become one of the most widely used techniques for palm detection and counting. Thus, their data effectively partly or entirely eliminates the need for physical counting\cite{ZiYanChen2020}. Finding the best automatic palm tree detection approach based-HRS image has attracted the attention of many industries and academics for end stakeholders' services. Existing detection methods for oil palms/tree crowns can be classified into three categories: (1) Traditional image processing methods; such as maximum local filter, image binarization, image segmentation, and template matching \cite{Quackenbush2017}, etc. (2) Traditional machine learning-based methods, generally stand in need to feature extraction, classifier training, image segmentation, and prediction, etc. \cite{Hung2021}. (3) Deep learning-based methods; have been tested in the RS area since 2014 and have attained good performance in various studies based on multispectral and hyperspectral images \cite{FabienHWagner2020}, including land cover mapping, object detection, semantic segmentation, \cite{Ghandorh}, \cite{Wadii}, etc.

\subsection{Aims and outline of the study}
This study is a systematic review of the decade, which flesh out the first literature review of RS and AI-based methods for palm tree detection as far as we are aware. This study is organised by: Section 2 briefly describes the methods used to perform the systematic review; Synthetic findings and results of the research questions are outlined in Section 3. We are discussing some critical issues based on the overall overview of this study in Section 4. Lastly, our conclusions and thoughts are presented. 

\section{Review scope and methodology}

This work aims to gain knowledge of existing studies that apply various artificial intelligence (i.e., machine learning and deep learning) and remote sensing techniques to the palm agriculture industry in terms of resolving issues. As far as the authors know, no attempt could be found in the available literature about "palm detection and counting and delineation" as a particular case, especially the date palm, because research on it is limited.
For this review, we followed the guidelines provided by \cite{yosra}, which was done through a systematic review of the peer-reviewed journal papers and Meta-Analysis (PRISMA) approach \cite{DMoher}. 

\subsection{Research Questions}
Two primary research questions (RQs) have been identified to address the purpose of this study. They were specifically selected to extract the latest technologies and interesting aspects of the ML and DL methods developed for detecting and delineating palm trees. The review is guided by the following RQs: \\\\
\textit{\textbf{RQ1.}How many studies were carried out about the application of ML and DL to serve the date palm and oil palm industry during the previous ten years, and which countries were actively involved in these research works?}\\
\textit{\textbf{RQ2.}What are the different methods used in Palm Tree detection?} \\
\textit{\textbf{RQ2.1.}What type of remotely sensed images are used for palm detection and counting?}\\
\textit{\textbf{RQ2.2.}What ML/DL methods have been widely used in the last 10 years to serve the palm?} 

\subsection{Search phase}
The first step in our systematic review is identifying the information sources. The research was conducted on a variety of academic databases, digital libraries, and search engines, both academic and open source. We used the following sources for our study;  Science Direct, Elsevier, Scopus, ACM digital library, IEEE Xplore, google-scholar, Springer, Sensors, Web of science, and Researchgate.
The second step is identifying search terms based on previous search questions to get a set of keywords.  
\subsubsection{Definition of the research string}
Basic Search has been completed automatically. The research started with "oil palm OR date palm" AND "machine learning". Articles were retrieved, and abstracts were read to identify synonyms for keywords. Research data were used to gain an overview of the studies.
Then, from basic research experience, a more complex research chain was built to focus on publications that primarily detect and count palm trees in combination with ML and DL patterns.
A final search string queried the journal databases was as follows;
(("oil AND palm AND tree" OR "Elaeis AND guineensis") OR ("date AND palm AND tree" OR "Phoenix AND dactylifera") AND ("machine AND learning" OR "deep AND learning" OR "artificial AND intelligence" OR "fully AND convolutional" OR "convolutional AND network") AND ("detection" OR "counting")).
The term "AND" was used for the exact combination of two strings and the term "OR" for flexible investigations. 
The defined search query consisted of three main parts;
the first part is aimed at relevant publications concerning palm trees ( e.g., oil or date trees). The second part was to identify publications that used AI methodologies (e.g., machine learning and deep learning), and the third part was focused on finding documents that have worked on tree detection and counting.

\subsubsection{Study Selection Criteria}
To refine the research results, we use a set of Inclusion Criteria (CI) and Exclusion Criteria (EC) to identify the appropriate documents. 
Studies that do not meet the EC are ignored; furthermore, a selection process (i.e., screening) is applied to select articles relevant to our context. \\

\textbf{Inclusion criteria}\\
\textit{IC1.} Studies released over the 2011-2021 period. \\
\textit{IC2.} Studies should address one or more of the keywords. \\
\textit{IC3.} Studies to be published/in the press at journals and conferences. \\
\textit{IC4.} Studies must include responses to the RQs. \\
\textit{IC5.} The search is conducted according to the title, abstract, and full text. \\

\textbf{Exclusion criteria}\\
\textit{EC1.} Duplicated papers \\
\textit{EC2.} Documents that do not appear in English. \\
\textit{EC3.} Documents in which the full text is messing \\
\textit{EC4.} Documents not directly linked to palm tree detection/counting within DL/ML methods. \\
\textit{EC5.} Papers do not directly employ the attention mechanism within DL/ML methods. \\
\textit{EC5.} The proposed study is not approved by the articles. \\
\textit{EC6.} Articles with an overall summary without clear input. \\
\textit{EC8.} Chapters, magazines, and newspaper articles. \\

\subsection{Conduct of the systematic review}
\begin{figure}
\centering
\includegraphics[width=14cm, height=12cm]{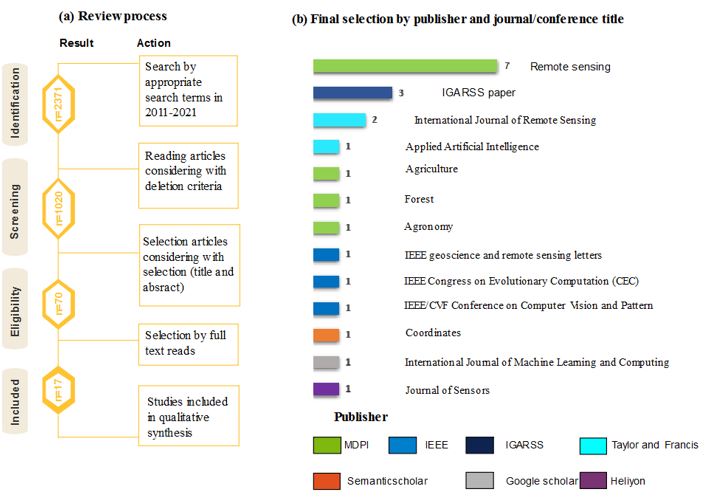}
\caption{(a) The review process (b) The distribution of selected papers by publisher.} 
\label{fig3}
\end{figure}

Articles relevant to this review were selected through the following key steps: identification, screening, eligibility, and inclusion, as described in the PRISMA flowchart. Fig \ref{fig3}.(a) Figure 2.(a) demonstrates the strategic-review process. Firstly, we obtained 2371 records identified through electronic databases and search engines.
This was reduced to 1020 without duplication. Secondly, the qualification criteria based on title and abstract were applied. This elimination round reduced our scores to 70; Other full-text eligibility criteria eventually resulted in 22 outcomes.
These 22 documents will be examined in detail, and their detailed findings and related discussions are presented for each RQ in the following sections.

\section{Results}

\begin{figure}
\centering
\includegraphics[width=14cm, height=7cm]{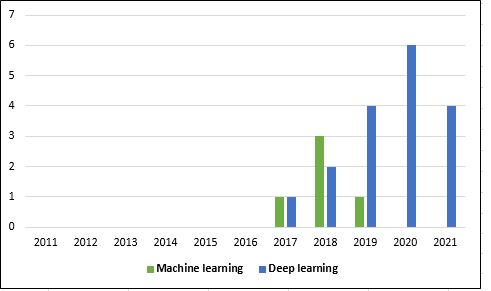}
\caption{The distribution of selected papers per year according to their main contribution field.} 
\label{fig4}
\end{figure}

\begin{figure}
\centering
\includegraphics[width=14cm, height=7cm]{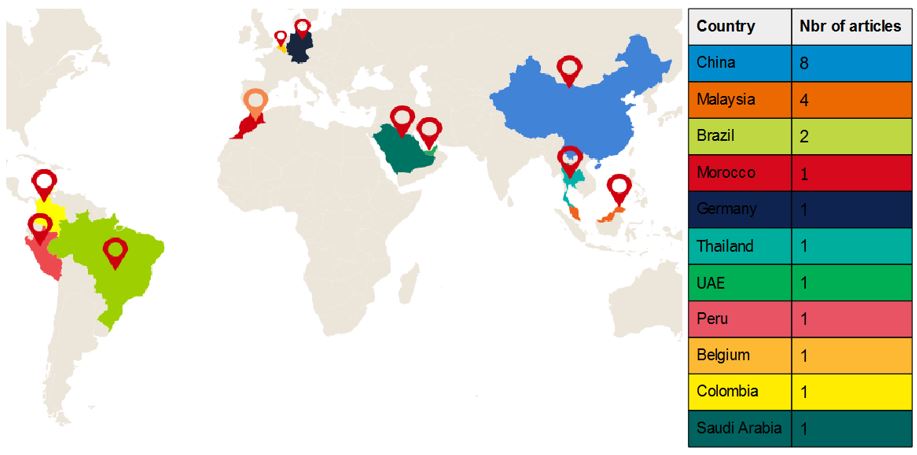}
\caption{Spatial distribution of author's study sites.} 
\label{fig5}
\end{figure}

\textit{Response to research question RQ1.\textbf{How many studies were carried out about the application of ML and DL to serve the date palm and oil palm industry during the previous ten years, and which countries were actively involved in these research works?} }\\
After executing iterative combinations of search strings defined within different databases mentioned in the previous, a final 22 potential papers have been selected for detailed review, see Fig.\ref{fig3}.(a). Despite the search period from 2011 to 2021, the appearance of NCs used to detect objects from RS images in the context of palm trees occurred in 2017 and expanded to reach a total of 22 papers in 2021, Fig.\ref{fig3} (a) illustrates the distribution of articles per editor, where most of the selected papers appear in RS journals with 7 papers followed by IGARSS conference with 3 papers, then by the international journals of RS by 2 papers. The annual distribution of selected papers is demonstrated in Fig. \ref{fig4} and the geographical distribution of author affiliations and study sites are shown in Fig. \ref{fig5}. Authors from China have contributed the most significant amount of papers 8 followed by Malaysia 4, Brazil 2, then equally distributed by the other mentioned countries. It is also worth pointing out that authors from Asia have contributed most of the studies than authors from North Africa.\\

\textit{Response to research question RQ2.1. \textbf{What type of RS imagery is used for palm detection and counting?}}\\

\begin{figure}[h]
\centering
\includegraphics[width=13cm, height=7cm]{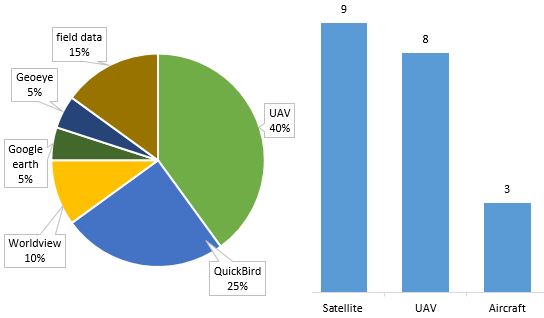}
\caption{Dataset and sources cited in the papers.} 
\label{fig6}
\end{figure}

\textit{RS Imaging:}  RS is known as a technique for detecting an object remotely to provide temporal and multi-modal data representing the surface of the Earth to obtain information \cite{Yosra1}. RS imaging techniques have shown great potential in palm plantation monitoring studies; this includes land cover classification \cite{Munirah}
, tree detection, counting, yield estimates, age estimation, pest and disease detection, etc., which contribute to effective productivity. Using RS as an alternative to the traditional method has broadened the scale and led many researchers to develop various techniques and methods. To decrease the cost of tree inventory \cite{Wulder}, automated tree crown detection based on high-resolution RS images has become one of the most appropriate and popular techniques to detect oil palm and counting. It is because counting trees on an image is much easier than counting on the floor, as they can be easily labeled on the image and can be easily re-counted for verification \cite{Mukhtar2014}.
Fig.\ref{fig6} summarizes the palm detection data, as presented in the 2011 to 2021 literature, which is categorized into satellite, aerial, and drone imagery.\\

\textit{Satellite image:} satellite RS has been used extensively for data collection over large areas for crop and agricultural applications, particularly plantation monitoring. 
In recent years, very high resolution (VHR) images with spatial resolutions below 1 m have been continuously acquired at the Earth's surface by VHR satellites (e.g., GeoEye-1 and Worldview-2) \cite{WeijiaLi2017}, \cite{Zouhayra}. The advent of such imaging opened the door for new applications such as automated detection and counting roofless buildings, object tracking, identification, etc. However, space and airborne platforms are not economical for small and medium-sized enterprises due to budget constraints.\\

\textit{Aircraft Imagery:}
the aircraft or aerial imagery generally refers to the image captured by an aircraft, helicopter, or manned aircraft. Usually these systems are equipped with sensors (i,e, infrared sensor, aerial laser scanner (ALS), electro-optical sensor, spectrometer, Light Detection and Ranging (LiDAR), multispectral and hyperspectral sensors (i.e, \cite{RamiAlRuzouq2018}, \cite{rizeei2018}, \cite{NurulainAbdMubin2019} ). In the reviewed studies, multiple flights were achieved at different altitudes (i.e., between 10m and 1000m) and speed to take photographs under diverse scenarios because they may impact the quality of the image acquisition process other than the spatial resolution of the camera used. On the other side, the sensor choice depends on the study's purpose.\\

\textit{Unmanned Aerial Vehicles (UAV):}
UAV computation technologies and sensors have been developed rapidly. The RS community increasingly adopts UAVs as new effective acquisition systems for RS applications in tree detection and delineation, besides VHR satellites. 
Many years ago, it was used only for military applications. However, due to technology improvement and cost reduction, they became an efficient solution for a large number of citizen applications according to customers' specific needs (i.e., monitoring, mapping, and counting palm trees) as they provide higher spatial resolution (up to 1 cm) and adjustable data acquisitions \cite{Mukhtar2014}. By contrast, UAV imagery presents additional challenges compared to VHR satellite imagery. Since improved spatial resolutions characterize them, traditional methods for analyzing RS images are not well suited.\\

\textit{Response to research question RQ2.2 \textbf{ Which ML/DL methods have been used extensively in the past 10 years to serve the palm?}}

The 22 selected studies guided the review by reading full text focusing on the following attributes; the applications studied, the data sets used, the CNN architectures and DL environments, and the results obtained. Thus, the CNN architecture reported for each article is the most effective architecture studied, particularly in ablation studies that compare several architectures. Results about the traditional machine learning-based methods and the deep learning-based methods are presented respectively in table \ref{tab1},  \ref{tab2} and \ref{tab3} (table \ref{tab2} continue).
%\begin{sidewaystable}
%\begin{landscape}
\begin{table}
\scriptsize
\centering
\caption{Overview of the main results in considered studies based on ML algorithms for palm tree detection}\label{tab1}
\begin{tabular}{|p{1cm}|p{3cm}|p{3cm}|p{4cm}|p{3cm}|}
\hline
\textbf{Refs} & \textbf{Method} & \textbf{Dataset} &  \textbf{Results} & \textbf{Objective(s)}  \\
\hline
\cite{BaharehKalantar2017}& SVM,TM,TM + OBA & UAV Images & OA: 87\% (TM+OBA), 71\% (TM) & Counting oil palm
inventory \\
\hline
\cite{YiranWang2018} & HOG-SVM,SVM  &UAV Images & OA: 99.21\%& Automatic detection and enumeration of individual oil pal trees  \\
\hline
\cite{RamiAlRuzouq2018} & RF, SVM, K-NN & Aerial remote sensing images& OA: 91.88\% and 87.03\%,& Date palm tree Detection and Mapping\\
\hline 
\cite{rizeei2018}& SVM of OBIA, FR + GIS &  WorldView-3 satellite, LiDAR airborne imagery.  & OA: 98,80\% for counting and 84.91\% for age estimation & Counting and age estimation of the oil palm tree.\\
\hline
\cite{taglecasapia2019} & Image processing + GIS functionalities, RF, SVMR, RP, k-NN. & UAV Images + ground data& Best OA: 85\% for RF with 0.82 kappa index &Identification and quantification of palm trees\\
\hline
\end{tabular}
\end{table}
%\end{landscape}
%\end{sidewaystable}

\begin{sidewaystable} 
\begin{landscape}
\centering

\caption{Overview of the main results in considered studies based on DL algorithms for palm tree detection }\label{tab2}
\begin{longtable}{|p{1cm}|p{4.50cm}|p{3cm}|p{3cm}|p{3cm}|p{3cm}|}
\hline

\textbf{Refs} & \textbf{Method} &\textbf{Backbone}& \textbf{Dataset} &  \textbf{Results} & \textbf{Objective(s)}  \\
\hline 
\cite{WeijiaLi2017}  & \textbf{The proposed method:} CNN, \textbf{Compared to:} ANN, TMPL, LMF & LeNet, ReLu & QuickBird satellite images & OA: 96\% & Detection and counting oil palm tree  \\
\hline
\cite{WeijiaLi2018} & \textbf{The proposed method:} TS-CNN \textbf{Compared to:}  single-stage CNN, SVM, RF and ANN/ Two stage method with different combinations of classifier&  LeNet, AlexNet and VGG-19 & QuickBird satellite images & F1-score: 94.99\% & Detecting oil palm trees\\  
\hline 
\cite{MacielZortea2018} & \textbf{The proposed method:} Two-CNN Slide + window technique & --- & UAV Images & OA: 91.2\% and 98.8\%. & Detecting oil palm trees\\
\hline 
\cite{NurulainAbdMubin2019}& \textbf{The proposed method:} Two different CNN + GIS
& LeNet, ReLu &WorldView-3 satellite images  & OA: 95.11\% and 92.96\% & Detection and counting young and mature oil palm trees  \\ 
\hline 
\cite{JuepengZheng2019} & \textbf{The proposed method:} Faster-RCNN + post-processing method \textbf{Compared to:} CNN, TS-CNN, Faster-RCNN  & VGG16 & QuickBird satellite images & Precision: 95.86\% , Recall: 95.30\% ,\par F1-score: 95.58\%  & Large scale Detection of oil palm trees\\
\hline
\cite{Xia2019} & \textbf{The proposed method:} VGG-SSD \textbf{Compared to:} faster-RCNN, YOLO-V3, Retina-net, Mobilenet-SSD & Resnet-50  &  UAV images & OA: 90.91\%   & Fast and robust detection of oil palm in large scale\\
\hline
\cite{Freudenberg2019} & \textbf{The proposed method:} U-Net  
Transfer learning \textbf{Compared to:} Alexnet with U-Net A and U- Net B simpler and optimized version  & Alexnet & Worldview-3&  A: Between 89\% and 92\%  & Large Scale Palm Tree Detection\\
\hline 
\cite{XinniLiu2020} & \textbf{The proposed method:} Faster-RCNN \textbf{Compared to:} SVM, ANN & VGG-16 + softmax & UAV Images & OA: 97.06\%, 96.58\%, and 97.79\% in 3 different sites & Automatic Detection of Oil Palm Tree\\ 

\hline 
\end{longtable}
\end{landscape}
\parbox{20cm}{---: Not mentioned}, {mAP: mean Average Precision}, {OA: Overall Accuracy}, {P: Precision}, {A: Accuracy} {R: Recall}
\end{sidewaystable}

\begin{sidewaystable}
\begin{landscape}
\centering
\caption{Overview of the main results in considered studies based on DL algorithms for palm tree detection } \label{tab3}
\begin{longtable}{|p{1cm}|p{4.50cm}|p{3cm}|p{3cm}|p{3cm}|p{3cm}|}
\hline
\textbf{Refs} & \textbf{Method} &\textbf{Backbone}& \textbf{Dataset} &  \textbf{Results} & \textbf{Objective(s)}  \\
\hline 
\cite{bonet2020}&\textbf {The proposed method:} CNN + Transfer learning + SVM & VGG-16 network without its
last layer + softmax& UAV Images & Accuracy: 97.5\%, Precision: 98.3\%.  & Detecting oil palm trees\\
\hline 
\cite{FabienHWagner2020}& \textbf{The proposed method:} U-Net &---& GeoEye-1 satellite images, & OA: 95.5\%, Canopy palm trees covered 6.4\% of the forest canopy& Mapping and Spatial Distribution Analysis of Canopy Palms; semantic segmentation\\
\hline 
\cite{MariaCulman2020}& \textbf{The proposed method:} RetinaNet + Transfer learning
 & Resnet-50 & Aerial remote sensing imagery and Palm Map & mAP: 0.861.& Detection of Palm Tree; Tree Inventory in large scale\\   
\hline 
\cite{ZiYanChen2020} & \textbf{The proposed method:} Fast R-CNN + SMF-driven DMS + data fusion (local maximum) filtering \textbf{Compared to:} Faster-RCNN, YoloV2 & VGG-16 & UAV Images & A: 99.8\%, 100\% and 91.4\%  in young, mature and mixed vegetation areas & Detection of Palm Tree in mixed study areas\\  
\hline 
\cite{WenzhaoWu2020} & \textbf{The proposed method:} CROPTD \textbf{Compared to:} Faster R-CNN, DA Faster R-CNN and Strong-Weak & ResNet-101 \cite{He2016} & QuickBird, Google Earth: satellite imagery & P : 90.06\%, R: 90.87\%, F1-score: 90.35\%  & Detection of oil palm \\ 
\hline 
\cite{HRhinane2021}  & \textbf{The proposed method:} DNN; U-Net + GIS & --- & Bing-images: satellite imagery & OA: 96.94\% & Detection and delineation of palm trees crown  \\
\hline 
\cite{AdelAnis2021}  & \textbf{The proposed method:} Faster R-CNN, YOLO (v3, v4) and EfficientDet & Resnet-50, Darknet-53, CSPDarknet-53, EfficientNet-B5 & UAV images &  mAP: 99\% of YOLO V4 and EfficientDet
FPS: 7.4  &  automated counting
and geolocation of palm trees   \\
\hline 
\cite{JuepengZheng2021} & \textbf{The proposed method:} MMD-DRCN & --- & QuickBird, Google Earth & P: 77.78\%, R: 89.47\% F1-score of 82.70\% & Detection of oil palm\\
\hline 
\cite{KanittaYarak2021}& \textbf{The proposed method:} Faster R-CNN& Resnet-50, VGG-16& UAV Images & P with Resnet-50: 96.34\%, P with VGG-16: 95.15\% & Oil Palm Tree Detection and Health Classification\\
\hline 
\end{longtable}
\end{landscape}
\parbox{20cm}{---: Not mentioned}, {mAP: mean Average Precision}, {OA: Overall Accuracy}, {P: Precision}, {A: Accuracy} {R: Recall}
\end{sidewaystable}

\section{Discussion} 

Unlike image-based methods, traditional ML-based methods do not need to manually select the tree planting area. They often outperform them in complex regions, thus having a better effect. Some of the literature reviewed based on RS detects not only tree crowns but also identifies tree species. To this end, ML-based \newpage methods can be used to evaluate various classifiers. In any case, using an unsupervised or supervised ML method would bet on the availability of data sets. In order to get good accuracy results, this method has some requirements for image datasets, as a wide range of spectral data can be delivered through high resolution, multispectral, hyperspectral, or thermal images. Unfortunately, these requirements bring issues to large-scale palm detection concerns somewhat. Besides, ML methods are deeply reliant on handcrafted features. A new handcrafted feature based on the latest data is necessary every time we have new data, which can be high-priced, and the process is tedious as it will be repeated.
The bottleneck of the handcrafted techniques demands an accurate features design during the feature extraction stage and fine-tuning of parameters in the training stage.\\

From the previous studies in (Table \ref{tab1}), the disadvantage is the detection's accuracy. Recently, numerous researchers have started extensively using deep learning methods because it shows excellent achievement in different applications. The aim of CNN-based methods for object detection techniques is to increase the efficiency of palm tree detection, as described in (Table \ref{tab2}). The main interest concerning CNN for object detection is that features can be learned automatically from images, which means that we do not need to have prior knowledge about the feature representation of the object or go through a complex handcrafted feature extraction process. It is kind of a fully data-driven scheme. The DL-based method was found to have outperformed other traditional tree detection methods. However, DL requires a massive volume of the sampling data set and complex data models to train it, which can be very costly. Furthermore, GPUs and hundreds of expensive machines would make costs go up. Additionally, with the introduction of transfer learning into DL models \cite{Freudenberg2019}, users can use a small dataset to refine an already formed network with countless images. This is quicker than training a CNN from scratch. Despite the fact that good tree detection results are obtained in small study areas (e.g., less than 5 km2), the actual DL-based tree detection methods are not appropriate within the large-scale study area because palm trees are very similar to other types of vegetation in different areas.\\

Each RS platform reviewed in this document has advantages and disadvantages in data acquisition. During the first decade, satellites and drones (i.e., UAVs) were unpopular, and investigations on tree detection and delineation were also quite limited. Satellite imagery offers better area coverage and provides many spectral data (e.g., panchromatic, multispectral, and hyperspectral data). Although, for aircraft and UAV images, the spectral and spatial data collected depend on the camera and the type of sensor used and on an area of limited coverage. However, thanks to the progress of cameras available in light and hyperspectral multispectral, with more up-to-date and cheaper. UAV is willing to be a better choice for personal, commercial uses, or small stakeholders than satellite imagery. The usage of UAVs and satellite images increased significantly between 2011 and 2021, and this trend is expected to increase with technological improvements. Overall, the use of SR as an alternative to the traditional approaches led many researchers to develop different techniques and ways to increase the inventory of palm trees. Whether it is satellite imagery, drones, conventional aerial photography, manned aircraft, or sensing technologies, the agricultural industry is changing dramatically.

\section{Conclusion}
In this study, we presented an investigation of current research on the application of AI in oil palm detection invoked by researchers in recent decades (2011–2021). A final 22 potential papers have been selected for detailed review. We categorized them into ML-based methods and DL-based methods. Overall, tree identification and delineation studies make a significant contribution to the inventory of palm plantations. However, there is still room for improvement and development, particularly in the methodologies used. Given the rising of RS availability, studies would focus more on the feasibility of the methodology used at a lower cost and with a higher quality outcome. The accuracy of tree detection results would depend not just on the algorithm but also on the data acquisition quality. The fusion of UAV and RS data could enhance tree detection accuracy and provide a more advanced approach in the future.

\end{document}